\documentstyle[psfig]{article}
\def\lsim{\:\raisebox{-0.5ex}{$\stackrel{\textstyle<}{\sim}$}\:}

\def\be{\begin{equation}}
\def\ee{\end{equation}}
\begin{document}

\title{Scaling in \boldmath{$\gamma^* \lowercase{p}$} total cross sections
and the generalized
vector dominance/color dipole picture\footnote{Supported by BMBF under
Contract 05HT9PBA2
\newline\indent
~~Presented at DIS 2001, Bologna, Italy, April 27 to May
1, 2001.}}

\author{D. Schildknecht\\[2mm] \em
Fakult{\"a}t f{\"u}r Physik, Universit\"at Bielefeld,\\
Universit{\"a}tsstra{\ss}e 25, D-33615 Bielefeld, Germany\\
E-mail: Dieter.Schildknecht@physik.uni-bielefeld.de
}

\maketitle

\begin{abstract}
The scaling in $\sigma_{\gamma^*p}$ cross sections (for $Q^2/W^2 << 1$)
in terms of the scaling variable $\eta = (Q^2 + m^2_0)/\Lambda^2 (W^2)$
is interpreted in the generalized vector dominance/color-dipole picture
(GVD/CDP). The quantity $\Lambda^2 (W^2)$ is identified as the average
gluon transverse momentum absorbed by the $q \bar q$ state,
$<\vec l^{~2}> = (1/6) \Lambda^2 (W^2)$. At any $Q^2$, for $W^2 \to \infty$,
the cross sections for virtual and real photons became universal,
$\sigma_{\gamma^*p}(W^2,Q^2)/\sigma_{\gamma p}(W^2) \to 1$.
\end{abstract}

Two important observations\cite{H1} were made on deep inelastic scattering
(DIS) at low values of the Bjorken scaling variable $x_{bj} \cong Q^2/W^2
<< 1$, since HERA started running in 1993:\\
i) The diffractive production of high-mass states (of masses
$M_X \lsim 30 GeV$) at an appreciable rate relative to the total
virtual-photon-proton cross section,
$\sigma_{\gamma^*p} (W^2,Q^2)$. The sphericity and
thrust analysis\cite{H1} of the diffractively produced states revealed
(approximate) agreement in shape with the final state found in $e^+e^-$
annihilation at $\sqrt s = M_X$. This observation of high-mass diffractive
production confirms the conceptual basis of generalized vector dominance
(GVD)\cite{Sakurai} that extends the role of the low-lying vector mesons in
photoproduction\cite{Stodolsky} to DIS at arbitrary $Q^2$, provided
$x_{bj} << 1$.\\
ii)An increase of $\sigma_{\gamma^*p}(W^2,Q^2)$ with increasing
energy considerably stronger\cite{Zeus} than the smooth ``soft-pomeron''
behavior known from photoproduction and hadron-hadron scattering.\\
We have recently shown\cite{Schi} that the data for total photon-proton cross
sections, including virtual {\it as well as real photons}, show a scaling
behavior.
In good approximation,
\be
\sigma_{\gamma^* p} (W^2, Q^2) = \sigma_{\gamma^* p} (\eta),\label{(1)}
\ee
with
\be
\eta = \frac{Q^2+m^2_0}{\Lambda^2(W^2)}.\label{(2)}
\ee
Compare Fig. 1. The scale $\Lambda^2(W^2)$, of dimension $GeV^2$,
turned out to be an increasing function of the $\gamma^* p$ energy,
$W^2$, and may be represented by a power law or a logarithmic function of
$W^2$,
\be
 \Lambda^2 (W^2) = \left\{ \begin{array} {r@{\quad~\quad}r}
c_1 (W^2+W^2_0)^{c_2},\\
c_1^\prime \ln (\frac{W^2}{W^{\prime 2}_0} + c^\prime_2).
\end{array}
\right. \label{(3)}
\ee
In a model-independent fit to the experimental data, the threshold
mass, $m^2_0 < m^2_\rho$, and the two parameters
$c_2 (c_2^\prime)$ and $W^2_0 (W^{\prime 2}_0)$ were found to be given by
$m^2_0  =  0.125 \pm 0.027 GeV^2,~
c_2  =  0.28 \pm 0.06,~
W^2_0  =  439 \pm 94 GeV^2$
with $\chi^2/ndf = 1.15$, and
$m^2_0 = 0.12 \pm 0.04 GeV^2,~
c_2^\prime  = 3.5 \pm 0.6,~
W_0^{\prime 2} = 1535 \pm 582 GeV^2,$
with $\chi^2/ndf = 1.18$. The overall normalization, $c_1 (c^\prime_1)$ in
(3) is irrelevant for the scaling behavior.

\begin{figure}[h]
\setlength{\unitlength}{1cm}
\begin{minipage}[t]{5.5cm}
\vspace*{1.4cm}
\begin{picture}(3.5,3.5)\psfig{file=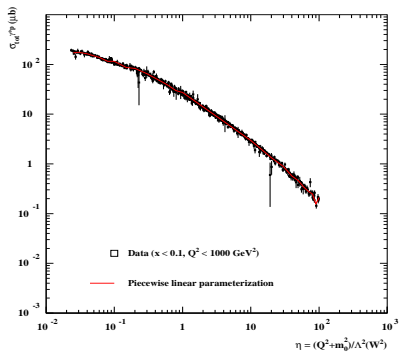,width=5.5cm,height=5.5cm}
\end{picture}\par
\end{minipage}\hfill
\begin{minipage}[t]{5.5cm}
\caption{The experimental data for $\sigma_{\gamma^*p}(W^2, Q^2)$ for
$x \simeq Q^2/W^2 \le 0.1$, including $Q^2 = 0$,
vs. the scaling variable $\eta = (Q^2 +
m^2_0)/\Lambda^2(W^2)$}
\end{minipage}
\vspace*{-0.5cm}
\end{figure}


For the interpretation of the scaling law (1) , we turn to the generalized
vector dominance/color-dipole picture (GVD/CDP)\cite{GVD,Schi}, of
deep-inelastic
scattering at low $x << 1$. It rests on $\gamma^*(q \bar q)$ transitions
from $e^+e^-$ annihilation, forward scattering of the $(q \bar q)$ states of
mass $M_{q \bar q}$ via (the generic structure of) two-gluon exchange\cite{7}
and
transition to spacelike $Q^2$ via propagators of the $(q \bar q)$ states
of mass $M_{q \bar q}$. In the transverse-position-space
representation\cite{Nikolaev}, we have
\begin{eqnarray}
\sigma_{\gamma^*p} (W^2, Q^2) & = &
\int dz \int d^2 r_\perp \vert \psi \vert^2
(r^2_\perp Q^2 z (1-z), Q^2 z (1-z), z) \cdot \nonumber \\
& \cdot &\sigma_{(q \bar q)p} (r^2_\perp, z(1-z),W^2).
\label{(4)}
\end{eqnarray}
We refer to ref.\cite{Nikolaev} for the explicit representation of the
square of
the photon wave function, $\vert \psi \vert^2$. The ansatz (4) for the
total cross section {\it must} be read in conjunction with the Fourier
representation of the color-dipole cross section,
\be
\sigma_{(q \bar q)p} (r^2_\perp, z (1-z),W^2)= \int d^2 l_\perp
\tilde \sigma_{(q \bar q)p} (\vec l^{~2}_\perp, z (1-z),W^2) \cdot
(1 - e^{- \vec l_\perp \cdot \vec r_\perp}).\label{(5)}
\ee
Upon insertion of (5) into (4), together with the Fourier
representation of the photon wave function, one indeed recovers\cite{GVD}
the expression for $\sigma_{\gamma^*p}$ that displays the
$x \to 0$ generic structure
of two-gluon exchange\footnote{It is precisely the identical
structure\cite{Nikolaev} that justifies the GVD/CDP (4), (5) from
QCD.}: The resulting expression for $\sigma_{\gamma^*p}$ is characterized
by the
difference of a diagonal and
an off-diagonal term with respect to the transverse momenta (or masses) of
the ingoing and outgoing $q \bar q$ states.

From (5), the color-dipole cross section, in the two limiting cases
of vanishing and infinite interquark separation, becomes, respectively,
\be
\sigma_{(q \bar q)p} (r^2_\perp, z (1-z),W^2) = \sigma^{(\infty)}
\cdot \left\{ \begin{array}{l@{\quad,\quad}l}
\frac{1}{4} r^2_\perp \langle \vec l^{~2} \rangle_{W^2,z}
& {\rm for}~ r^2_\perp
\to 0,\\
1 & {\rm for}~ r^2_\perp \to \infty.
\end{array}
\right. \label{(6)}
\ee
The proportionality to $r^2_\perp$ for small interquark separation is known
as ``color transparency''\cite{Nikolaev}. For large interquark separation,
the color-dipole cross section should behave as an ordinary hadronic one.
Accordingly,
\be
\sigma^{(\infty)} = \pi \int dl^2_\perp \tilde \sigma (l^2_\perp,
z (1-z),W^2)\label{(7)}
\ee
must be independent of the configuration variable $z$ and has to fulfill
the restrictions from unitarity on its energy dependence. The average gluon
transverse momentum $\langle \vec l^{~2}\rangle_{W^2,z}$ in (6), is
defined by
\be
\langle \vec l^{~2} \rangle_{W^2,z} = \frac{\int d\vec l^{~2}_\perp
\vec l^{~2}_\perp
\tilde \sigma_{(q \bar q)p} (\vec l^{~2}_\perp, z(1-z),W^2)}
{\int d \vec l^{~2}_\perp
\tilde \sigma_{(q \bar q)p} (\vec l^{~2}_\perp, z(1-z),W^2)}.\label{(8)}
\ee
Replacing the integration variable $r^2_\perp$ in (4) by the
dimensionless
variable
\be
u \equiv r^2_\perp \Lambda^2 (W^2) z (1-z),\label{(9)}
\ee
the photon wave function becomes a function $\vert \psi \vert^2 (u
\frac{Q^2}{\Lambda^2}, \frac{Q^2}{\Lambda^2},z)$. The requirement of scaling
(1), in particular for $Q^2 >> m^2_0$, then implies that the
color-dipole cross sections be a function of $u$,
\be
\sigma_{(q \bar q)p} (r^2_\perp, z (1-z), W^2) = \sigma_{(q \bar q)p} (u).
\label{(10)}
\ee
Taking into account (6), we find
\be
\langle \vec l^{~2}\rangle_{W^2,z} = \Lambda^2 (W^2) z (1-z),\label{(11)}
\ee
and upon averaging over $z$,
\be
\langle \vec l^{~2} \rangle_{W^2} = \frac{1}{6} \Lambda^2 (W^2).\label{(12)}
\ee
The quantity $\Lambda^2 (W^2)$ in the scaling variable (2)
is accordingly identified as the average gluon transverse momentum, apart
from the factor 1/6 due to the averaging over $z$.

Inserting $\langle \vec l^{~2} \rangle_{W^2,z}$ from (11) into (6),
we have
\be
\sigma_{q \bar q p} = \sigma^{(\infty)} \cdot
\left\{ \begin{array}{l@{\quad,\quad}l}
\frac{1}{4} r^2_\perp \Lambda^2 (W^2) z (1-z) & {\rm for}~\Lambda^2 \cdot
r^2_\perp \to 0,\\
1 & {\rm for}~\Lambda^2 \cdot r^2_\perp \to \infty.
\end{array} \right.\label{(13)}
\ee
The dependence of the photon wave function in (4) on $r^2_\perp \cdot
Q^2$ requires small $r_\perp$ at large $Q^2$, in order to develop appreciable
strength; for large $Q^2$, the $r^2_\perp \to 0$ behavior in (13),
with its associated strong $W$ dependence, becomes relevant until, finally,
for sufficiently large $W$,
the soft $W$ dependence of $\sigma^{(\infty)}$ will be reached.

Thus, by interpreting the empirically established scaling, $\sigma_{\gamma^*,p}
= \sigma_{\gamma^*p}(\eta)$, in the GVD/CDP, we have obtained the dependence
of the color-dipole cross section on the dimensionless variable $u$ in
(10) and, consequently, with (13), qualitatively,
the dependence on $\eta$ shown in fig. 1. Conversely, assuming a functional
form for the color-dipole cross section according to (10), one
recovers the scaling behavior (1).

In\cite{Schi}, we have shown that approximating the distribution in the gluon
momentum transfer by its average value, (11),
\be
\tilde \sigma_{(q \bar q)p} = \sigma^{(\infty)} \frac{1}{\pi} \delta
(\vec l^{~2}_\perp - \Lambda^2 (W^2) z (1-z)),\label{(14)}
\ee
allows one to analytically evaluate the expression for $\sigma_{\gamma^*p}$
in (4) in momentum space. The threshold mass $m_0 \lsim m_\rho$ enters
via the lower limit of the integration over the masses of the ingoing
and outgoing $q \bar q$ states. For details we refer to\cite{Schi}, and only
note the approximate result
\be
\sigma_{\gamma^*p}(\eta) \simeq \frac{2 \alpha}{3 \pi} \sigma^{(\infty)}
\cdot \left\{ \begin{array}{l@{\quad,\quad}l}
\ln (1 \vert \eta) &{\rm for}~\eta \to \eta_{\rm min} =
\frac{m^2_0}{\Lambda^2(W^2)},\\
1 \vert 2 \eta &{\rm for}~\eta >> 1.
\end{array} \right. \label{(15)}
\ee
Note that for any fixed value of $Q^2$, with $W^2 \to \infty$,
the soft logarithmic
dependence as a function of $\eta^{-1}$ is reached. We arrive
at the important conclusion that in the $W^2 \to \infty$ limit virtual and
real photons become equivalent,
\be
\lim_{{W^2 \to \infty} \atop {Q^2 {\rm fixed}}} \frac{\sigma_{\gamma^*p}
(W^2,Q^2)}{\sigma_{\gamma p} (W^2)} = 1.\label{(16)}
\ee
Even though convergence towards unity is extremely slow, such that it
may be difficult to ever be verified experimentally, the universality of real
and virtual photons contained in (16) is remarkable. It is an outgrowth
of the HERA results which are consistent with the  scaling law (1) with
$\eta$ from (2) and
$\Lambda^2 (W^2)$ from (3). Note that the alternative of
$\Lambda^2 = const$ that implies Bjorken scaling of the structure function
$F_2 \sim Q^2 \sigma_{\gamma^*p}$ for sufficiently large $Q^2$, leads to
a result entirely different from (16),
\be
\lim_{{W^2 \to \infty} \atop {Q^2 {\rm fixed}}} \frac{\sigma_{\gamma^*p}
(W^2,Q^2)}{\sigma_{\gamma p} (W^2)} = \frac{\Lambda^2}{2 Q^2 \ln
\frac{\Lambda^2}{m^2_0}},~~({\rm assuming}~ \Lambda = const.),\label{(17)}
\ee
i.e. a suppression of the virtual-photon cross section by a power of $Q^2$.
\begin{figure}[h]
\setlength{\unitlength}{1cm}
\begin{minipage}[t]{5.5cm}
\begin{picture}(3.5,3.5)\psfig{file=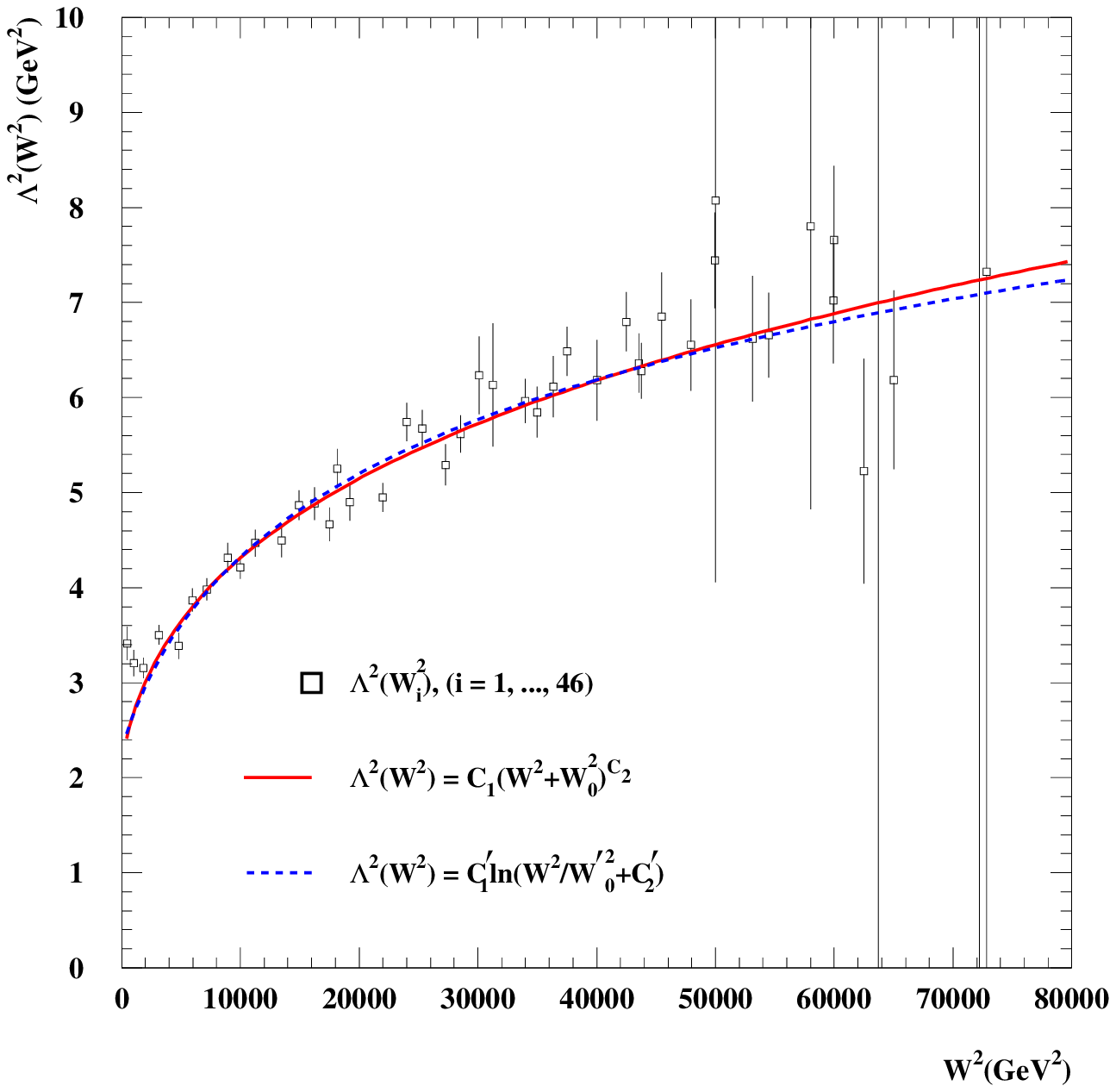,width=5.5cm,height=5.5cm}
\end{picture}\par
\caption{The dependence of $\Lambda^2$ on $W^2$, as determined by a fit
of the GVD/CDP predictions for $\sigma_{\gamma^*p}$ to the experimental
data}
\end{minipage}\hfill
\begin{minipage}[t]{5.5cm}
\begin{picture}(5.5,5.5)\psfig{file=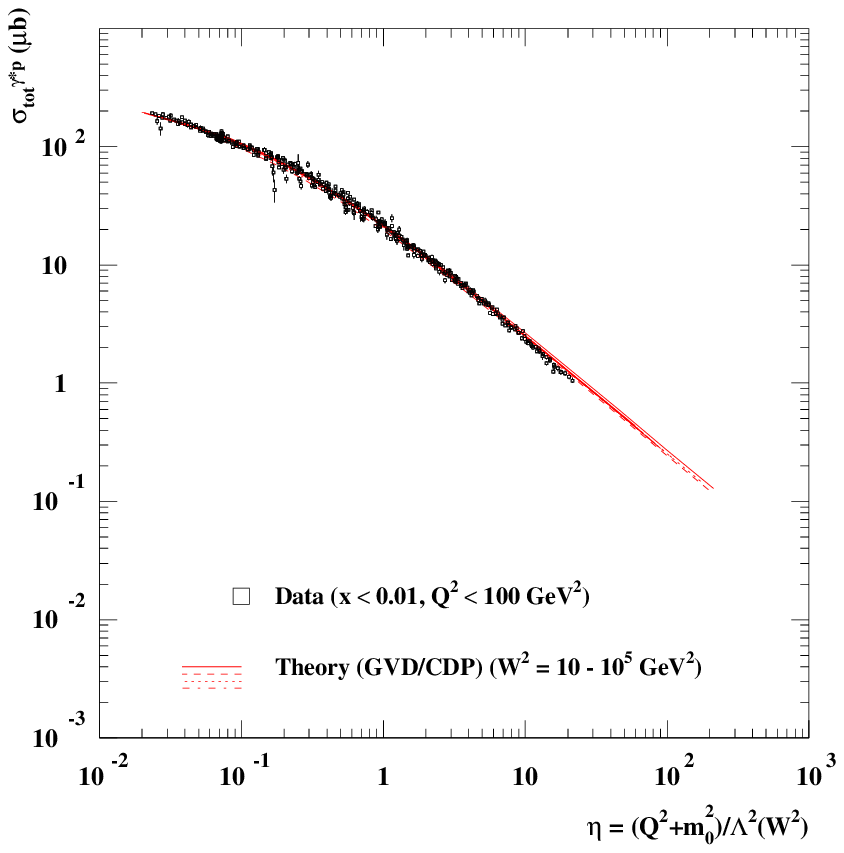,width=5.5cm,height=5.5cm}
\end{picture}\par
\caption{The GVD/CDP scaling curve for $\sigma_{\gamma^*p}$ compared with
the experimental data for $x < 0.01$}
\end{minipage}
\end{figure}

In Fig. 2, we show $\Lambda^2(W^2)$ as obtained from the fit\cite{Schi} of
$\sigma_{\gamma^*p}$ to the experimental data. The figure shows the result
of fits based on the power law and the logarithm in (3), as well as
the results of a pointlike fit, $\Lambda^2(W^2_i)$. Using (12), one
finds that the average gluon transverse momentum increases from
$<\vec l^{~2}> \simeq 0.5 GeV^2$ to $<\vec l^{~2}> \simeq 1.25 GeV^2$ for
$W$ from $W \simeq 30 GeV$ to $W \simeq 300 GeV$. In Fig. 3, we show the
agreement between theory and experiment for $\sigma_{\gamma^*p}$ as a
function of $\eta$. For
further details we refer to ref.\cite{Schi}.

In summary, we have shown that the HERA data on DIS
in the low-$x$ diffraction region find a natural
interpretation in the GVD/CDP. The scale $\Lambda^2 (W^2)$ entering the
scaling variable $\eta$, was found to be proportional to the average gluon
transverse momentum absorbed by the incoming (outgoing)
$q \bar q$ state in the
virtual-forward-Compton amplitude. The cross sections for real and virtual
photons on protons become identical in the limit of infinite energy.

\section*{Acknowledgments}
It is a pleasure to thank G. Cvetic, B. Surrow and M. Tentyukov for a
fruitful collaboration that led to the results reported here.

\end{document}